\documentclass[traditabstract]{aa}
\bibpunct{(}{)}{;}{a}{}{,} 
\usepackage{graphicx}
\usepackage[varg]{txfonts}
\usepackage{color}
\usepackage[utf8]{inputenc}
\usepackage{hyperref}
\begin{document}

\title{
Origin and evolution of NiI and FeI in the coma of the interstellar comet 3I/ATLAS throughout its trajectory}
\author{Damien Hutsem\'ekers\inst{1},
        Jean Manfroid\inst{1},
        Cyrielle Opitom\inst{2},
        Emmanu\"el Jehin\inst{1},
        Aravind Krishnakumar\inst{1},
        Fernando Massa Fernandes\inst{3},
        Michele Bannister\inst{4},
        Dennis Bodewits\inst{5},
        Rosemary Dorsey\inst{6},
        Fiorangela La Forgia\inst{7},
        Brian Murphy\inst{2}
        }
\institute{
    STAR, Institut d'Astrophysique et de G\'eophysique, Universit\'e de Li\`ege, All\'ee du 6 Ao\^ut 19c, 4000 Li\`ege, Belgium
    \and
    Institute for Astronomy, University of Edinburgh, Royal Observatory, Edinburgh EH9 3HJ, United Kingdom
    \and 
    Institute of Condensed Matter and Nanosciences, Universit\'e Catholique de Louvain, Chemin du Cyclotron 2, 1348 Louvain-la-Neuve, Belgium
    \and
    School of Physical and Chemical Sciences - Te Kura Mat\={u}, University of Canterbury, Private Bag 4800, Christchurch 8140, New Zealand
    \and
    Auburn University, Department of Physics, Auburn, AL 36849, USA
    \and
    Department of Physics, University of Helsinki, P.O. Box 64, 00014 Helsinki, Finland
    \and
    Department of Physics and Astronomy, University of Padova, vicolo Osservatorio 3, 35020 Padova, Italy
    }

\date{Received ; accepted: }
\titlerunning{NiI and FeI in 3I/ATLAS} 
\authorrunning{D. Hutsem\'ekers et al.}
\abstract{
We present high-resolution UVES+VLT observations of neutral nickel (NiI) and iron (FeI) atoms in the coma of the interstellar comet 3I/ATLAS taken after perihelion at heliocentric distances ($r_h$) ranging from 1.88 to 4.35 au. Metal emission was strong shortly after perihelion and persisted at large heliocentric distances. Spatial profiles indicate a short-scale source for both NiI and FeI. At all heliocentric distances, NiI dominates the metal budget. At $r_h \sim 2$~au the total metal production rate was found to be at least an order of magnitude higher than that of typical Solar System comets. Post-perihelion production rates exhibit clear asymmetry compared to the pre-perihelion behavior: production rates are higher after perihelion and decline more gradually with $r_h$, with the difference being stronger for FeI. The NiI/FeI abundance ratio, initially anomalously high before perihelion, evolved toward values comparable to Solar System comets near 2~au, and shows a weaker $r_h$ dependence after perihelion. To interpret these results, we revisited and extended the carbonyl hypothesis in which FeI and NiI are produced by the rapid photodissociation of Fe(CO)$_5$ and Ni(CO)$_4$ vaporized from the nucleus. Using sublimation laws parameterized with simple temperature profiles, $T(r_h)$, we fitted the full pre- and post-perihelion datasets and demonstrated that the direct sublimation of carbonyls can reproduce the observed rates and the high NiI/FeI abundance ratio. This ratio is controlled by the relative volatilities and active fractions of Ni(CO)$_4$ and Fe(CO)$_5$. The temperature profiles needed to reproduce the observations were found to be shallower than the equilibrium $T \propto r_h^{-1/2}$ relation. The full dataset could also be reproduced using the temperature profiles extracted from a thermal model of the nucleus, but only with significant empirical modifications to the model. In particular, a significant reduction in the active areas through the perihelion was required. An additional transient heat source ($T \sim$ 120~K) was also necessary. This heat source, possibly related to the amorphous-crystalline ice transition, could explain the early NiI excess before perihelion. Within the framework of this modified thermal model, carbonyl sublimation occurs below the surface of the nucleus. Although carbonyls provide a self-consistent explanation for the presence of metal atoms in cometary comae at large heliocentric distances, the direct detection of carbonyls in comets remains elusive. Alternative mechanisms, such as release from superheated metallic nanoparticles, cannot be ruled out.
}
\maketitle
\nolinenumbers
%
%
%

\section{Introduction}
\label{sec:intro}

\begin{table*}[]
\caption{Observing circumstances.}
\label{tab:obs1}
\centering
\begin{tabular}{lccccccc}
\hline\hline
 Date       & $r_h$ &  $\dot{r_h}$ & $\Delta$ & $\dot{\Delta}$ & Settings & $w_B / w_R$ & $h_B / h_R$ \\
 yyyy-mm-dd & au    &  km s$^{-1}$ & au        & km s$^{-1}$    &             & \arcsec & \arcsec \\
\hline 
2025-12-04$^{\ast}$    & 1.88 & 43.3 & 1.88 & -17.1   &      580   &    -/0.6 &    -/11.5   \\
2025-12-06$^{\dagger}$  & 1.93 & 44.5 & 1.86 & -15.5   &  348+580   &  0.6/0.6 &  9.5/11.5   \\
2025-12-10            & 2.04 & 46.5 & 1.83 & -11.6   &  437+860   &  0.6/0.6 &  9.5/11.0   \\
2025-12-15            & 2.18 & 48.6 & 1.80 &  -5.8   &  348+580   &  0.6/0.6 &  9.5/11.5   \\
2025-12-21            & 2.35 & 50.6 & 1.80 &   2.8   &  348+580   &  0.6/0.6 &  9.5/11.5   \\
2025-12-21            & 2.35 & 50.6 & 1.80 &   2.8   &  437+860   &  0.6/0.6 &  9.5/11.0   \\
2025-12-26            & 2.50 & 51.9 & 1.82 &  11.3   &  348+580   &  0.6/0.6 &  9.5/11.5   \\
2026-01-11            & 2.99 & 54.7 & 2.05 &  38.9   &  348+580   &  1.2/1.2 &  9.5/11.5   \\
2026-01-19            & 3.24 & 55.6 & 2.26 &  50.9   &  348+580   &  1.2/1.2 &  9.5/11.5   \\
2026-01-27            & 3.50 & 56.3 & 2.52 &  61.4   &  348+580   &  1.8/1.8 &  9.5/11.5   \\
2026-02-07            & 3.85 & 57.0 & 2.94 &  71.9   &  348+580   &  1.8/1.8 &  9.5/11.5   \\
2026-02-22            & 4.35 & 57.6 & 3.61 &  81.9   &  348+580   &  1.8/1.8 &  9.5/11.5   \\
\hline
\end{tabular}
\tablefoot{ $r_h$ and $\Delta$ are the heliocentric and geocentric distances of the comet. $\dot{r_h}$ and $\dot{\Delta}$ are the corresponding radial velocities. $w_B$, $w_R$, $h_B$, and $h_R$ refer to the blue/red slit width and height, respectively. ($^{\ast}$): No observation with the setting 348 due to a technical problem. ($^{\dagger}$): This observation was done through clouds.}
\end{table*}

Atomic emission lines of iron and nickel were first identified in the spectra of two Sun-grazing comets: the Great Comet of 1882 \citep{Copeland1882} and comet C/1965 S1 (Ikeya-Seki; \citealt{Dufay1965,Thackeray1966,Preston1967,Slaughter1969}). In the cometary material, iron and nickel have been found in refractory dust grains, metallic grains, and sulfides \citep{Zolensky2006}. Close to the Sun, the temperature of the comet is high enough to vaporize the refractory grains, releasing iron and nickel atoms that fluoresce when exposed to the solar radiation. In comet Ikeya-Seki, which was observed at a heliocentric distance of about 0.1 au, the NiI/FeI abundance ratio derived from the atomic emission lines was found to be comparable to that of the Sun and meteorites \citep{Preston1967,Arpigny1979}, and consistent with the Ni/Fe abundance ratio directly measured in dust grains collected from the comae of comets P/Halley \citep{Jessberger1988} and 81P/Wild2 \citep{Flynn2006}.

\citet{Manfroid2021} reported the presence of FeI and NiI emission lines in the high-resolution spectra of approximately 20 comets observed with the Very Large Telescope (VLT) at the European Southern Observatory (ESO) using the UV-Visual Echelle Spectrograph (UVES). The comets were observed at heliocentric distances ($r_h$) ranging from 0.68 to 3.25~au. At these distances, the temperature of the comets is not high enough to vaporize the refractory grains containing iron and nickel atoms. Furthermore, the NiI/FeI abundance ratio was found to be one order of magnitude higher than the nickel-to-iron abundance ratio measured in the Sun, meteorites, cometary dust grains, and the Sun-grazing comet Ikeya-Seki. Based on these observations, \citet{Manfroid2021} suggested that the NiI and FeI atoms could be released into the gas phase either from superheated metallic nanoparticles, as proposed for atomic sodium \citep{Ip1998}, or from a short-lived volatile parent, such as organometallic compounds or metal carbonyls, which have been proposed as constituents of the cometary, meteoritic, and interstellar material \citep{Huebner1970,Bloch1980,Klotz1996}. With sublimation temperatures between those of CO$_2$ and H$_2$O, thus much lower than the temperatures needed to vaporize metallic grains and sulfides, the Fe(CO)$_5$ and Ni(CO)$_4$ carbonyls would be able to release NiI and FeI atoms at large heliocentric distances. Furthermore, the overabundance of NiI relative to FeI can be interpreted as a direct consequence of the higher sublimation rate of Ni(CO)$_4$. Further discussions can be found in \citet{Manfroid2021}, \citet{Hutsemekers2021}, \citet{Bromley2021}, and \citet{Rahatgaonkar2025}.

The two known interstellar comets, 2I/Borisov (hereafter 2I) and 3I/ATLAS (hereafter 3I), originate from planetary systems beyond our Solar System. Thus, these comets serendipitously provided a rare opportunity to study the process by which iron and nickel atoms are released in comets that have traveled greater distances and experienced different temperature conditions than those in our Solar System. In comet 2I, FeI and NiI were found with a NiI/FeI abundance ratio similar to that of the Solar System comets \citep{Guzik2021,Opitom2021}. In comet 3I, NiI emission lines were identified above the dust continuum at a heliocentric distance of 3.78~au, before CN, which was detected at 3.65~au \citep{Rahatgaonkar2025}. The first FeI emission lines appeared at 2.64~au \citep[][hereafter Paper~I]{Hutsemekers2026}. During the initial stages of its gaseous activity, comet 3I  first exhibited an unusually high NiI/FeI abundance ratio. As the heliocentric distance decreased, the comet developed a rich spectrum of NiI and FeI emission lines and the NiI/FeI ratio became indistinguishable from those observed in Solar System comets and in comet 2I. Nevertheless, comet~3I remained exceptional due to its high total production rate of NiI and FeI, which was at least one order of magnitude higher than that of other comets (Paper I).

Comet~3I also appeared exceptional in other respects: it had very high CO$_2$/H$_2$O and CH$_3$OH/HCN pre-perihelion abundance ratios \citep{Cordiner2025,Biver2026,Roth2026a}. After perihelion, at heliocentric distances greater than 2~au, the apparent chemical composition changed drastically. It shifted from a H$_2$O-dominated coma near perihelion to a CO-dominated coma with a high volatile-to-water abundance ratio \citep{Biver2026,Lisse2026,Roth2026c}. Furthermore, significant isotopic anomalies were found, pointing to a comet formed in a cold environment at a large distance from a low-metallicity central star \citep{Cordiner2026,Opitom2026,Roth2026b,Salazar2026}.  For an exhaustive account of the many observations of comet 3I, see \citet{Frincke2026}.

In this paper, we present post-perihelion measurements of the NiI and FeI production rates using UVES spectra secured with the VLT. These observations were obtained at regular intervals, for comet heliocentric distances ranging from 1.88 au to 4.35 au. They complement the pre-perihelion measurements discussed in Paper I. Building on the scenario developed in Paper I, we interpret the full dataset assuming that the NiI and FeI atoms were released through the sublimation of Ni(CO)$_4$ and Fe(CO)$_5$ carbonyls.

\section{Observations, data reduction, and production rate  measurements}
\label{sec:obs}

\begin{table*}[]
\caption{FeI and NiI production rates, and their ratio.}
\label{tab:data}
\centering
\begin{tabular}{lccccccccccc}
\hline\hline
 Date & Setting & $r_h$ & n$_{\rm lines}$  & log$_{10}$ Q(FeI) & log$_{10}$ Q(NiI) & log$_{10}$ [Q(NiI)/Q(FeI)]  \\
 yyyy-mm-dd & & au     &  FeI / NiI  &   s$^{-1}$     &   s$^{-1}$ &        &     \\
\hline 

2025-12-04$^{a}$       & 580 & 1.88 &    8/0   &  24.31  $\pm$ 0.04  &  -   &  -    \\
2025-12-06$^{\dagger}$  & 348 & 1.93 &   55/37  &  23.93  $\pm$ 0.03  & 24.09 $\pm$ 0.05  &  0.16 $\pm$ 0.06   \\
2025-12-06$^{a,\dagger}$ & 580 & 1.93 &   10/0   &  24.10  $\pm$ 0.04  & -    &  -   \\
2025-12-10$^{b}$       & 437 & 2.04 &   43/3   &  23.95 $\pm$  0.03 &  24.17 $\pm$ 0.06  &  0.22 $\pm$ 0.06   \\
2025-12-15            & 348 & 2.18 &   54/37  &  23.78 $\pm$ 0.03  &  24.05 $\pm$ 0.05  &  0.26 $\pm$ 0.06   \\
2025-12-21            & 348 & 2.35 &   57/37  &  23.51 $\pm$ 0.03  &  23.93 $\pm$ 0.05  &  0.43 $\pm$ 0.06   \\
2025-12-21$^{b}$       & 437 & 2.35 &   42/3   &  23.43 $\pm$ 0.03  &  23.83 $\pm$ 0.05  &  0.40 $\pm$ 0.06   \\
2025-12-26            & 348 & 2.50 &   54/35  &  23.29 $\pm$ 0.03  &  23.82 $\pm$ 0.05  &  0.53 $\pm$ 0.06   \\
2026-01-11            & 348 & 2.99 &   24/31  &  22.66 $\pm$ 0.05  &  23.45 $\pm$ 0.05  &  0.79 $\pm$ 0.07   \\
2026-01-19            & 348 & 3.24 &   12/23  &  22.37 $\pm$ 0.09  &  23.17 $\pm$ 0.06  &  0.80 $\pm$ 0.11   \\
2026-01-27            & 348 & 3.50 &    3/17  &  22.10 $\pm$ 0.17  &  22.94 $\pm$ 0.06  &  0.84 $\pm$ 0.18   \\
2026-02-07            & 348 & 3.85 &    0/7   &   -                &  22.51 $\pm$ 0.08   &  -   \\
2026-02-22            & 348 & 4.35  &   0/0   &   -                &  -   &  -   \\
\hline
\end{tabular}
\tablefoot{{The uncertainties in the production rates account for those affecting the measured intensities and the atomic data used in the model. Additionally, when considering the absolute values of the production rates (not their ratio), the following systematic uncertainties should be considered: approximately 0.1~dex due to potential errors in absolute flux calibration and approximately 0.3~dex due to unknown ejection velocity.}    (a): FeI lines are measured at $\lambda > 5160$~\AA; no useful NiI lines in that wavelength range. (b) Only three NiI lines can be measured in this spectral range. ($^{\dagger}$): The production rates are underestimated due to atmospheric extinction (clouds); the ratio is not affected.}
\end{table*}

Post-perihelion observations of comet 3I were carried out from December 4, 2025, to February 22, 2026, using the VLT with UVES. During that period of time, the comet's heliocentric distance increased from 1.88 to 4.35 au. Metal lines were detected in the UVES settings 348, 437, and 580, which cover the spectral ranges 3100-3900~\AA,  3730-4990~\AA, and 4760-6840~\AA, respectively. When the comet was at a heliocentric distance of less than 2.5 au, a 0.6\arcsec\  wide slit was used, delivering a resolving power of about 65000.  As the comet dimmed, the slit was progressively opened to 1.2 and 1.8\arcsec, providing resolving powers of approximately 40000 and 35000, respectively.  The observing circumstances are summarized in Table~\ref{tab:obs1}.

Data reduction and spectral extraction were performed as described in Paper I for the pre-perihelion observations. Figure~\ref{fig:spec_nife} shows the spectrum of comet 3I in the spectral range 3100-5500~\AA, five weeks after perihelion. The spectrum is clearly dominated by numerous NiI and FeI emission lines, some of them as bright as the CN band that is observed around 3875~\AA. As the comet moved away from the Sun, the metal lines gradually disappeared. The FeI lines could not be detected in the February 7 spectrum ($r_h$ = 3.85 au) and in subsequent spectra.  In our final spectrum obtained on February 22 ($r_h$ = 4.35 au), two faint NiI lines were only barely detected. 

As in Paper~I, we measured the intensities ($I_{\rm obs}$) of the unblended FeI and NiI emission lines for each epoch. To derive the FeI and NiI column densities, these intensities were compared to theoretical values ($I_{\rm mod}$) computed using a dedicated multilevel fluorescence model\footnote{As explained in Appendix~\ref{app:3levels}, the three-level model, also described in \citet{Manfroid2021}, should not be used to derive absolute production rates, because it can overestimate the column densities by up to one order of magnitude compared to the multilevel model.} presented in \citet{Manfroid2021}. This model considers the intricate structure of the NiI and FeI energy levels and their various transitions. Using a high-resolution solar spectrum and accounting for the relative velocity between the comet and the Sun, the model calculates the actual solar flux experienced by each atomic transition. The column densities derived for each line from the ratio $I_{\rm obs}/I_{\rm mod}$ were then averaged for each epoch. The error of the mean value was evaluated using the standard deviation of the measurements after rejecting outliers. Since forbidden NiI infrared lines were observed in comet 3I \citep{Belyakov2026,Roth2026c}, we modified the selection of the transitions used in \citet{Manfroid2021} to include these low-probability transitions. Thus, we considered all transitions from the Atomic Line List \citep{Vanhoof2018} where the energy of the lower level was less than 20000 cm$^{-1}$, resulting in 1025 NiI transitions. Increasing the threshold on the energy of the lower level has no effect on the results. Similarly, 5385 transitions were considered for FeI. At the longest infrared wavelengths, the solar spectrum was extrapolated using a blackbody spectrum. For FeI, the new transition selection has a negligible effect on the computed column densities. For NiI, however, a better correspondence between the observed and modeled line intensities is achieved by taking the infrared transitions into account. Furthermore, the resulting column densities of NiI are slightly higher (on average 0.1~dex) and show a lower dispersion than when these transitions are not considered.

Since the error in the average column densities is evaluated from the dispersion of the measurements obtained for individual lines, it includes random errors from the intensity measurements and the atomic data. To further investigate the uncertainties from the model, we compared our measurements to those obtained using Steven Bromley's publicly available code \citep{Bromley2021,Bromley2026}, which utilizes atomic data from the National Institute of Standards and Technology (NIST) Atomic Spectra Database \citep{Kramida2026}. Overall, the agreement is very good, as seen in Fig.~\ref{fig:sbcompa}. The comparison nevertheless indicates a residual dispersion of 0.03~dex for FeI and 0.05~dex for NiI, mostly due to the different sources of atomic data. These uncertainties are added in quadrature to the error budget. 

\begin{figure}[]
\centering
\resizebox{0.95\hsize}{!}{\includegraphics*{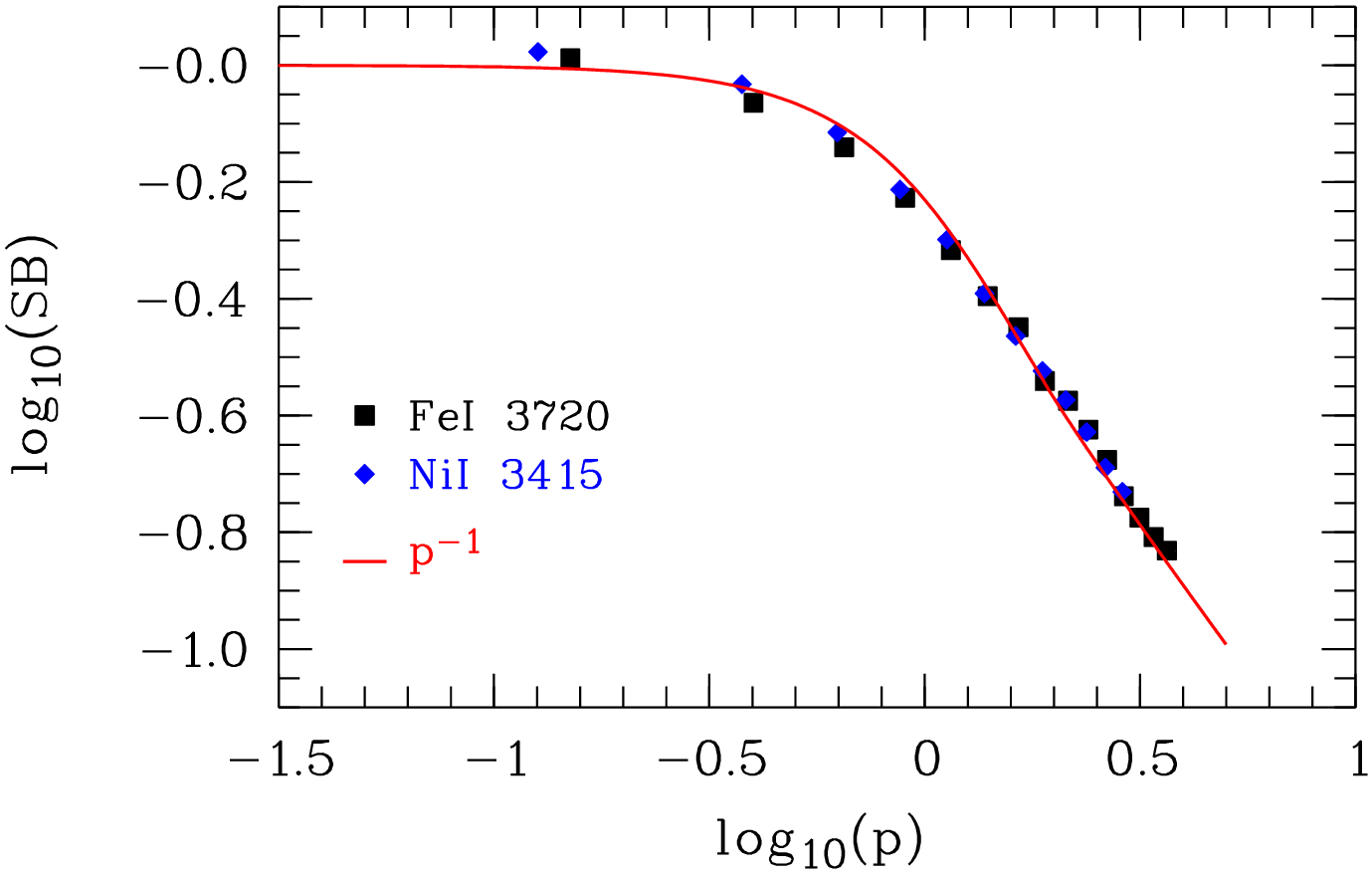}}
\caption{Post-perihelion spatial profiles of two bright NiI and FeI lines observed on December 21. The surface brightness (SB), normalized to 1 at the photocenter, is plotted as a function of the projected nucleocentric distance ($p$) in arcseconds. The red line represents SB $\propto p^{-1}$, convolved with a 1.5\arcsec\ full width at half maximum Gaussian to account for the seeing and tracking imperfections.}
\label{fig:rp}
\end{figure}

The NiI and FeI emission lines in comet 3I are characterized by a short spatial extent. Their surface brightness (SB) decreases rapidly following the relation SB $\propto p^{-1}$, where $p$ is the projected distance to the nucleus (Fig.~\ref{fig:rp}). This behavior was also observed pre-perihelion (Paper~I\footnote{The line used in Fig.~3 of Paper I was in fact NiI $\lambda$ 3415.}) and in Solar System comets \citep{Manfroid2021,Hutsemekers2021}. Using this relation and assuming a standard ejection velocity of $v$ = 0.85 $r_h^{-0.5}$ km s$^{-1}$ \citep{Cochran1993}, the production rates were computed from the column densities as detailed in \citet{Manfroid2021}. The post-perihelion production rates derived for NiI and FeI in comet 3I are given in Table~\ref{tab:data}. Most measurements were obtained with the setting 348 that contains the largest number of NiI and FeI lines. Production rates measured using another setting on the same night (setting 580 on December 6, and setting 437 on December 21) exhibit systematic differences of 20-40\%, most likely due to the uncertainties in the absolute flux calibration. Therefore, to create the plots shown in the next section, we slightly rescaled the production rates measured with settings 437 and 580 to align them with the rates measured with setting 348, using measurements obtained the same nights. To the production rates measured with settings 580 and 437, we added -0.17$\pm$0.03~dex and +0.08$\pm$0.02~dex, respectively. The Q(NiI)/Q(FeI) ratios are independent of this correction.

Several measurements of HCN, H$_2$O, CO$_2$, and CO obtained when comet 3I was at heliocentric distances of about 2 au, indicate ejection velocities ranging from 0.2 to 0.5  km~s$^{-1}$ \citep{Biver2026,Cordiner2026,Coulson2026,Roth2026a}. These velocities are approximately half of those given by the relation $v$ = 0.85 $r_h^{-0.5}$ km s$^{-1}$. Since production rates scale with the ejection velocity as Q $\propto N \times v$, where $N$ is the column density averaged over the aperture, velocities a factor of two lower would lead to production rates a factor of two lower. Without a better knowledge of the ejection velocity of the NiI and FeI atoms, and how it depends on $r_h$, we used the standard relation, bearing in mind that the production rates in Tables~\ref{tab:data} and~\ref{tab:data1} could be overestimated by a factor of two. Even if they were lower by 0.3~dex, the NiI and FeI production rates in comet 3I would still be higher than those of most comets in the Solar System (see Sect.~\ref{sec:results}). A shift in absolute Q values has no impact on our results, which are mostly related to the relative variation of the production rates with $r_h$.

Finally, although the differences are small, we revised the pre-perihelion production rates using the same atomic transition list that was used to compute the post-perihelion production rates, that is, including the infrared transitions. These values are provided in Table~\ref{tab:data1} and are used in the subsequent plots.

In Appendix~\ref{app:datacompa} we compare our measurements of the NiI and FeI production rates in comet 3I with recently published values. Unless explicitly noted, all published production rates were computed with an ejection velocity $v$ = 0.8 - 0.85 $r_h^{-0.5}$ km~s$^{-1}$,  allowing for a direct comparison with our values. Although the measurements were based on various observational techniques, spectral resolutions, methods of analysis, etc, they agree within the uncertainties, with the exception of the production rates reported by \citet{Roth2026c}, which are higher.

\section{NiI and FeI post-perihelion evolution}
\label{sec:results}

\begin{figure}[]
\centering
\resizebox{0.95\hsize}{!}{\includegraphics*{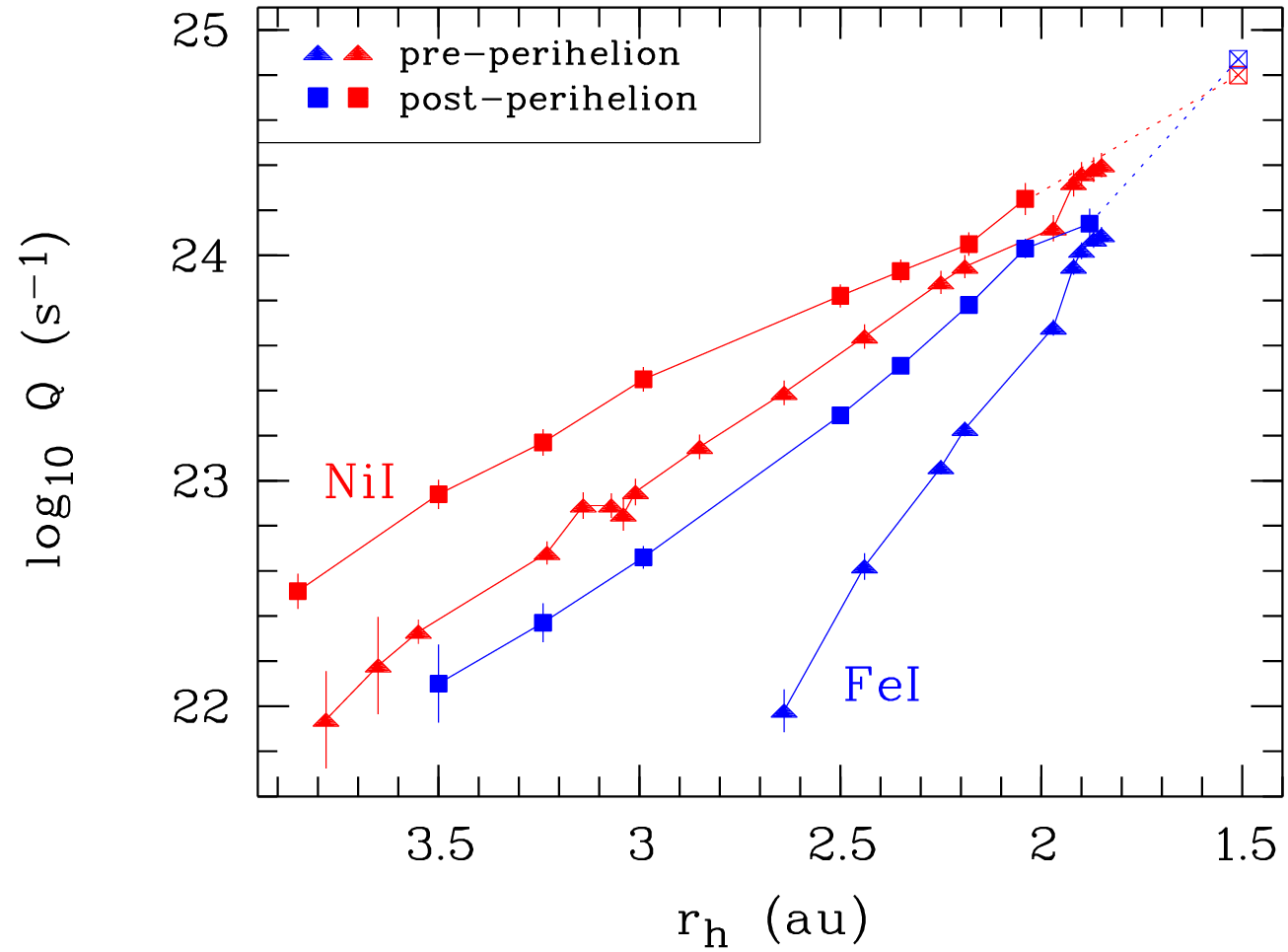}}
\caption{Production rates of NiI and FeI in comet 3I as a function of the heliocentric distance, $r_h$. Pre- (post-) perihelion measurements are from Table~\ref{tab:data1} (Table~\ref{tab:data}). The values obtained with settings 437 and 580 are slightly rescaled, as explained in Sect.~\ref{sec:obs}. The production rates measured on December 6 are not plotted. The open squares represent the measurements obtained by \citet{Hoogendam2026} shortly after perihelion (see Appendix~\ref{app:datacompa}). In addition to possible shifts of approximately 0.1~dex due to errors in absolute flux calibration, the production rates could be affected by a systematic shift of approximately $-$0.3~dex if the ejection velocity were half as high as assumed (see text). }
\label{fig:q01}
\end{figure}

\begin{figure}[]
\centering
\resizebox{0.95\hsize}{!}{\includegraphics*{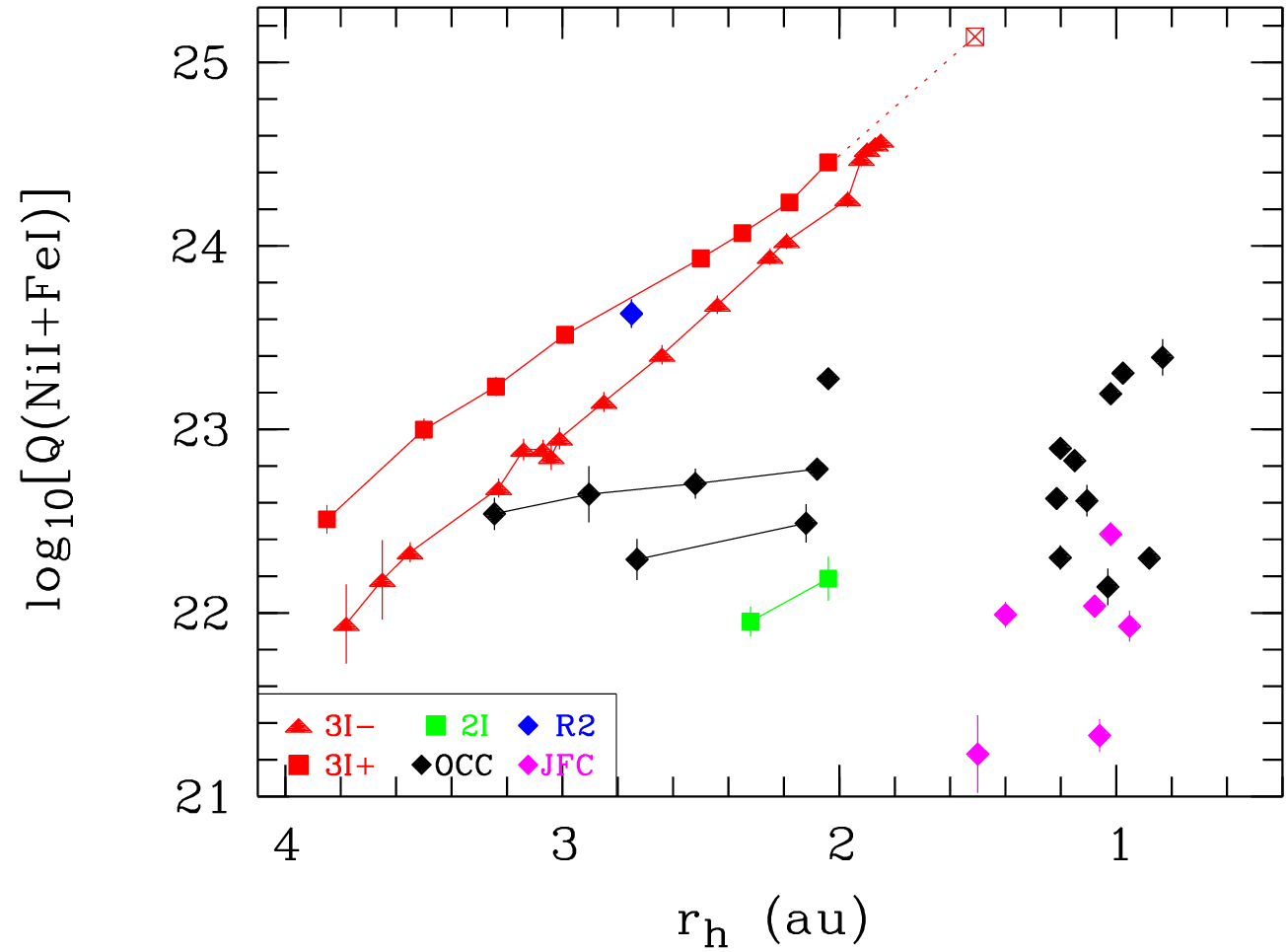}}
\caption{Q(NiI)+Q(FeI) as a function of the heliocentric distance ($r_h$) for the interstellar comets 2I and 3I, together with Solar System comets. Measurements of comet 3I are from Tables~\ref{tab:data1} and~\ref{tab:data}, adjusted as in Fig.~\ref{fig:q01}.  The open square represents the  measurement of \citet{Hoogendam2026}. The other data are described in Paper~I. In addition to comet 3I, measurements obtained at various heliocentric distances are shown for comets 2I (two values, post-perihelion), C/2009 P1 (four values, pre-perihelion), and C/2017 K2 (two values, pre-perihelion). For each of these comets, the values are connected by solid lines.}
\label{fig:c01}
\end{figure}

Figure~\ref{fig:q01} shows the evolution of the NiI and FeI production rates in comet 3I before and after perihelion. A significant asymmetry is evident for both species.  Post-perihelion production rates are higher than pre-perihelion rates and decrease more slowly as the heliocentric distance increases. This difference is more pronounced for FeI than for NiI. At approximately 2.5~au,  the FeI production rate is one order of magnitude higher post-perihelion than pre-perihelion. Such an asymmetry is not observed for other species analyzed using the same instrumentation \citep{Aravind2026}. Compared to other comets (Fig.~\ref{fig:c01}), the production rate of metals in comet 3I remains exceptionally high post-perihelion, with a steep variation as a function of the heliocentric distance.

\begin{figure}[]
\centering
\resizebox{0.95\hsize}{!}{\includegraphics*{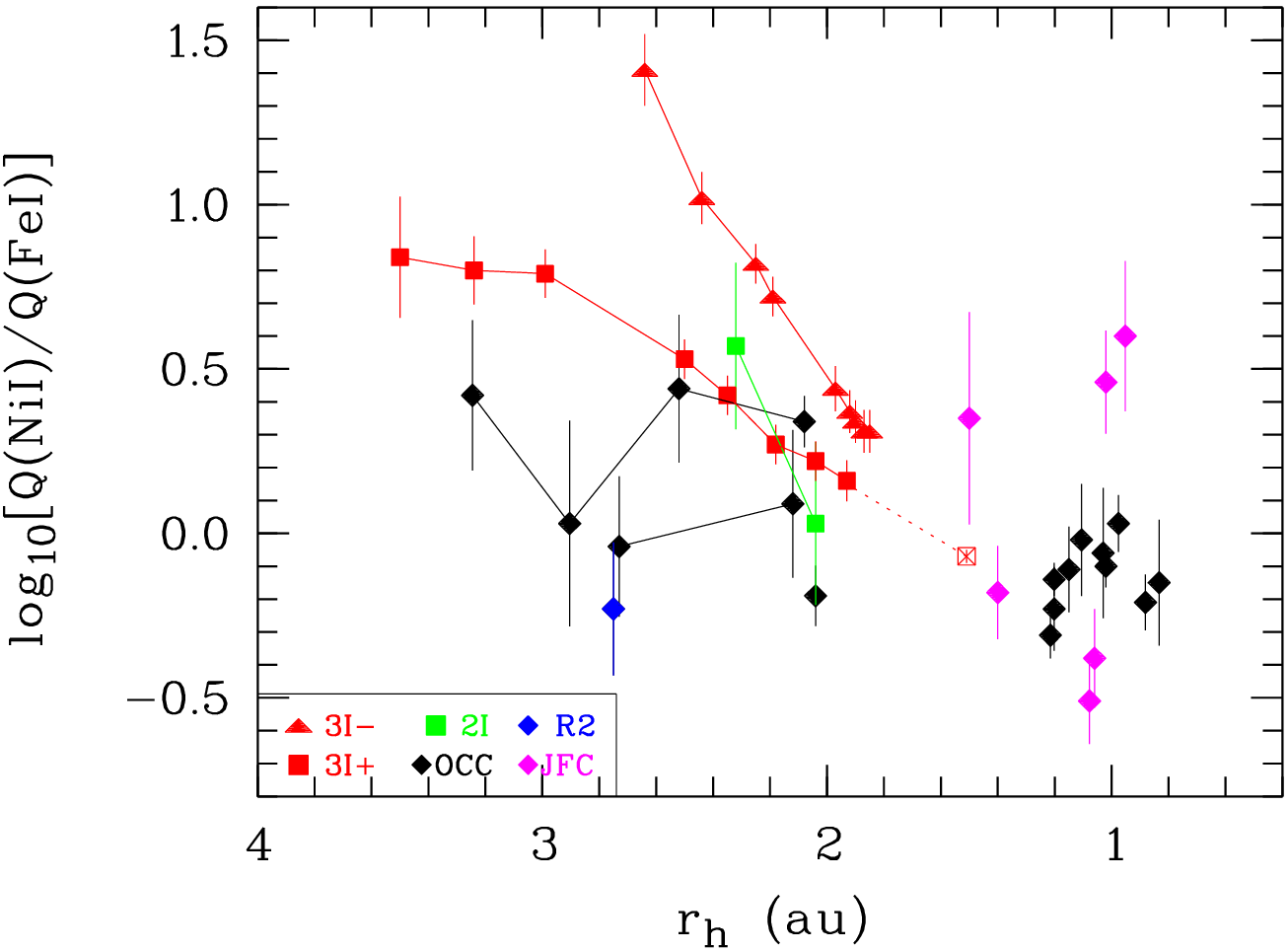}}
\caption{Same as Fig.~\ref{fig:c01} but for the ratio Q(NiI)/Q(FeI).  The average Q(NiI)/Q(FeI) ratio of Solar System comets is approximately zero on a log scale \citep{Manfroid2021}. The solar Ni/Fe abundance ratio is much smaller and equal, in logarithm,  to $-1.25 \pm 0.04$ \citep{Asplund2009}.}
\label{fig:c05}
\end{figure}

As shown in Fig.~\ref{fig:c05}, the Q(NiI)/Q(FeI) abundance ratio in comet 3I is comparable to that observed in Solar System comets, at heliocentric distances around and less than 2~au. This ratio varies more slowly with the heliocentric distance after perihelion than before perihelion, reaching values that are higher, yet still comparable, to those observed in Solar System comets. On the contrary, the Q(NiI)/Q(FeI) ratio was exceptionally high in the initial pre-perihelion observations. In our observations, Q(NiI) is always higher than Q(FeI).  However, \citet{Hoogendam2026} reported Q(FeI) $>$ Q(NiI) closer to perihelion, at $r_h$ = 1.51~au (see Appendix \ref{app:datacompa}), in agreement with the trend seen at higher $r_h$ (Fig.~\ref{fig:c05}), and the predictions from the carbonyl sublimation model (Paper~I and Fig.~\ref{fig:fitall}).

\begin{figure}[]
\centering
\resizebox{0.95\hsize}{!}{\includegraphics*{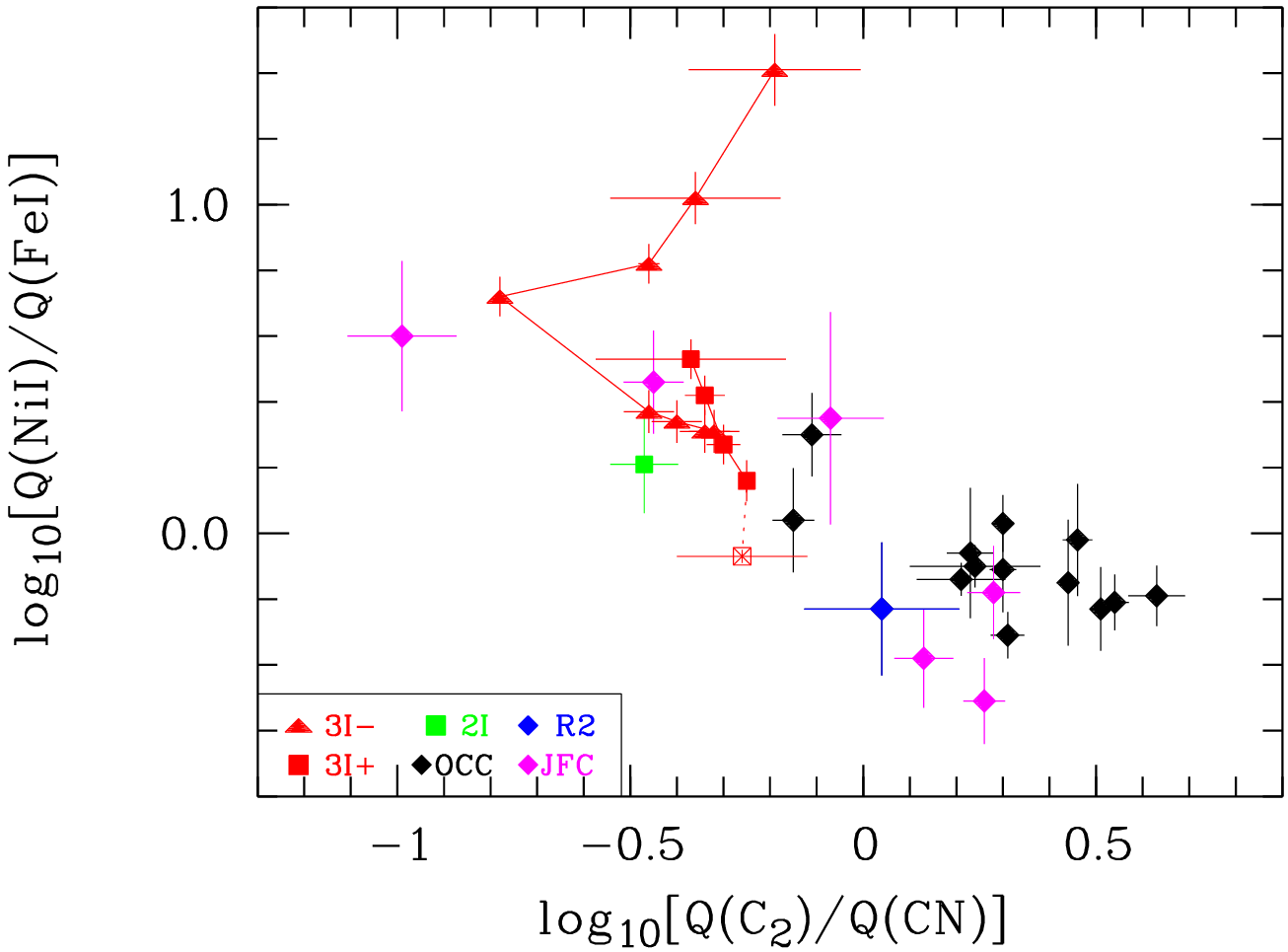}}
\caption{Q(NiI)/Q(FeI) as a function of Q(C$_2$)/Q(CN). Pre-perihelion Q(C$_2$)/Q(CN) abundance ratios are from Paper~I and post-perihelion values from \citet{Aravind2026}. The open square represents the measurements obtained by \citet{Hoogendam2026} for comet 3I.}
\label{fig:c33}
\end{figure}

Finally, the post-perihelion Q(NiI)/Q(FeI) ratios align with the correlation between Q(NiI)/Q(FeI) and Q(C$_2$)/Q(CN) reported in \citet{Hutsemekers2021}. They align more closely than the pre-perihelion values, which exhibit extremely high Q(NiI)/Q(FeI) values at $r_h$ > 2.4~au (see Fig.~\ref{fig:c33}).

As previously noted in Paper~I, comet 2I is more similar to comets of the Solar System than to comet 3I with regard to metal production. Only comet C/2016 R2 exhibits a total metal production rate comparable to that of 3I (Fig.~\ref{fig:c01}), although this is a tentative connection as comet R2 was only measured at a single heliocentric distance (2.8~au). However, the metal production in comet C/2016 R2 is dominated by FeI, unlike comet 3I where it is dominated by NiI, except close to perihelion (Fig.~\ref{fig:c05}).

Thus, comet 3I is clearly exceptional in terms of metal production compared to the interstellar comet 2I and typical Solar System comets, as already suggested from the pre-perihelion observations \citep{Rahatgaonkar2025,Hutsemekers2026}. The origin of this difference is unclear. However, unusual isotopic ratios \citep{Cordiner2026,Opitom2026,Roth2026b,Salazar2026} point to a comet formed in a cold environment. This also points to a low metallicity star, which at first glance could look incompatible with the high NiI and FeI abundances. Nevertheless, the high abundance of CO in comet~3I \citep{Lisse2026,Roth2026c,Zhao2026} may indicate that it contains an unusually large reservoir of volatile material, in particular metal-bearing volatiles. This material may have been preserved because the comet formed in a cold environment. In any case, metal-bearing volatiles would represent only a fraction of the total metal content, since most Fe and Ni are expected to remain locked in refractory phases such as metallic grains, silicates, and sulfides, which can be efficiently released only very close to the Sun.

\section{Origin of NiI and FeI: The carbonyl scenario}
\label{sec:models}

In Paper~I, we investigated a scenario in which the FeI and NiI atoms are released from Fe(CO)$_5$ and Ni(CO)$_4$ carbonyls.  We showed that the pre-perihelion evolution of the FeI and NiI production rates could be explained by this scenario, particularly the high and variable NiI/FeI abundance ratio that was observed in comet 3I. Here, we revisit this scenario in more detail taking into account the post-perihelion observations.

\subsection{The model}

 We write the production rates of FeI and NiI as a function of the sublimation rates ($Z$) of Fe(CO)$_5$ and Ni(CO)$_4$:
\begin{align*}
{\rm Q(FeI)} \;  & = a_{\rm Fe} \; Z_{{\rm Fe(CO)}_5} ,\\
{\rm Q(NiI)} \;  & = a_{\rm Ni}  \; Z_{{\rm Ni(CO)}_4} .
\end{align*}
The parameters $a_{\rm Fe}$ and $a_{\rm Ni}$ represent the effective (active) areas that release Fe(CO)$_5$ and Ni(CO)$_4$ into the gas phase due to direct sublimation from the icy nucleus. These relations implicitly assume that the photodissociation of Fe(CO)$_5$ and Ni(CO)$_4$ into atomic FeI and NiI is very fast and similar for the two species \citep{Bromley2021}. The sublimation rate from the surface of pure ice into vacuum can be expressed as
\begin{equation}
\label{eq:z}
Z_x (T) = \frac{P_{v,x} (T)}{\sqrt{2 \pi m_x k T}} 
,\end{equation}
where $T$ is the ice temperature and $m_x$ the mass of the species~$x$ \citep{Delsemme1982}. $P_{v,x}$, the vapor pressure (in dyn cm$^{-2}$), is given by the relation
\begin{equation}
\label{eq:p}
\log_{10} P_{v,x} (T) = -A/T \, + \, B \;.
\end{equation}
The constants ($A,B$) for Fe(CO)$_5$ and Ni(CO)$_4$ were obtained from \cite{Gilbert1974} and \citet{Stull1947}: ($A,B$) = (2097, 11.62) for  Fe(CO)$_5$ and ($A,B$) = (1534, 10.87) for Ni(CO)$_4$.  Finally, the dependence of the temperature $T$ on the heliocentric distance $r_h$ was parametrized through the relation
\begin{equation}
T = T_{\rm p} \; (r_h/r_p)^{\gamma},
\end{equation}
where $T_{\rm p}$ is the temperature at the perihelion, $r_p$ = 1.357 au.  $T_{\rm p}$ and $\gamma$ are parameters to determine. Since the blackbody equilibrium temperature of the cometary surface is approximately $T \simeq 280 \; r_h^{-1/2}$, where $r_h$ is in au \citep{Manfroid2021, Guzik2021,Puzia2025}, we expect $T_{\rm p} \simeq$ 240~K and $\gamma \simeq -0.5$ at equilibrium.

\begin{figure*}[]
\centering
\resizebox{0.95\hsize}{!}{\includegraphics*{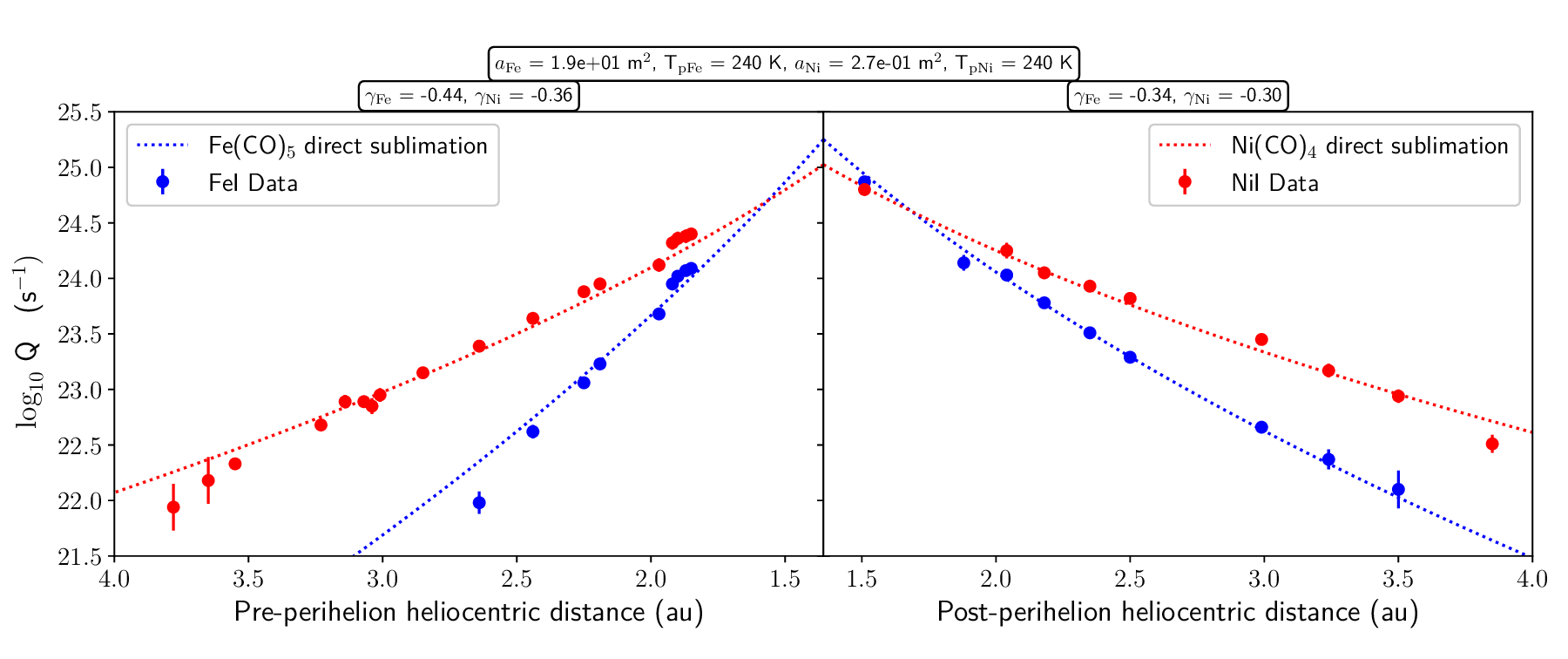}}
\caption{Production rates of NiI and FeI as a function of heliocentric distance. The dataset is the same as that shown in Fig.~\ref{fig:q01}. Post-perihelion measurements at $r_h$ = 1.51~au are from \citet{Hoogendam2026}. The model (dotted lines) is based on the direct sublimation of the Fe(CO)$_5$ and Ni(CO)$_4$ carbonyls, assuming free temperature profiles (see text). The corresponding fit parameters are given above each panel. }
\label{fig:fitall}
\end{figure*}

The key assumption of this scenario is that the production of gaseous atomic iron and nickel primarily depends on the direct sublimation of the Fe(CO)$_5$ and Ni(CO)$_4$ carbonyls from the icy nucleus of comet 3I. Based on a long-term monitoring of comet C/1995 O1 (Hale-Bopp), \citet{Biver1997,Biver2002} found that the production rates of minor species in the coma were not related to their sublimation temperatures, suggesting that minor species were released through desorption from major cometary ices (i.e., CO, CO$_2$, or H$_2$O), rather than direct sublimation. On the contrary, when comet 3I approached perihelion, the onset and production rates of the nickel and iron atoms were clearly in line with the sublimation temperatures of Ni(CO)$_4$ and Fe(CO)$_5$, which range from 74 to 82~K and 97 to 108~K, respectively \citep{Manfroid2021}. This sequence is reversed outbound (Fig.~\ref{fig:q01}), as expected from direct sublimation. Furthermore, desorption of gaseous carbonyls from CO, CO$_2$, or H$_2$O raises several issues with respect to the observations. First, desorption from CO is unlikely due to the much lower sublimation temperature ($\sim$ 25~K) of this species. As found by \citet{Rubin2023}, species of a given volatility (defined by the sublimation temperature) rarely desorb alongside species of higher volatility. Additionally, if carbonyls were released from CO, they would likely be released in their solid phase and would still require direct sublimation. This two-step process would result in a distributed source of gaseous carbonyls and metals, contradicting the observed short spatial scales. A priori, desorption from CO$_2$ and H$_2$O, with sublimation temperatures of approximately 72~K and 152~K, respectively, could produce gaseous Ni(CO)$_4$ and Fe(CO)$_5$ at heliocentric distances of less than about 3.5~au. In this case, one would expect the $r_h$-dependence of the NiI and FeI production rates to be comparable to that of CO$_2$ or H$_2$O, as observed for several minor species in comet Hale-Bopp \citep{Biver1997,Biver2002}. In comet 3I, the $r_h$-dependence of the FeI production rate, and to a lesser extent of the NiI production rate, significantly differs from the $r_h$-dependence of OH, a proxy for H$_2$O \citep{Aravind2026}. Furthermore, none of the species studied by \citet{Aravind2026} show the strong pre/post perihelion asymmetry of the NiI and FeI production rates (Fig.~\ref{fig:q01}). Finally, the simultaneous release of Ni(CO)$_4$ and Fe(CO)$_5$ in the gaseous phase through desorption would result in an abundance ratio NiI/FeI independent of $r_h$, which is contrary to the observations (Fig.~\ref{fig:c05}). One could argue that the  Ni(CO)$_4$ over Fe(CO)$_5$ abundance ratio in the ice phase depends on the carbonyl location on or below the nucleus surface, such that desorption could produce a varying NiI/FeI ratio in the gas phase. However, in this case, the continuing process of removing nucleus layers after the perihelion would not explain why the outbound NiI/FeI variation follows the reverse inbound sequence. This sequence is naturally explained by the direct sublimation of Ni(CO)$_4$ and Fe(CO)$_5$. Thus, the hypothesis that carbonyls enter the gaseous phase through direct sublimation from the icy phase is a reasonable starting point for the proposed scenario, at least for comet 3I. However, we cannot exclude other mechanisms as the processes by which carbonyls could form in cometary ices are basically unknown.

\subsection{Fitting the data}

We fitted the NiI and FeI pre- and post-perihelion data simultaneously, ensuring continuity of the temperatures, active areas, and production rates through the perihelion ($r_h$ = 1.357~au). Eight parameters were estimated: $T_{\rm pFe}$, $T_{\rm pNi}$, $\gamma_{\rm Fe pre}$, $\gamma_{\rm Fe post}$, $\gamma_{\rm Ni pre}$, $\gamma_{\rm Ni post}$, $a_{\rm Fe}$, and $a_{\rm Ni}$. The temperatures were upper-bounded by the equilibrium temperature $T_{\rm p} =$ 240~K. The dataset is the same as that shown in Fig.~\ref{fig:q01}, including the measurement obtained near perihelion by \citet{Hoogendam2026}. The fit is shown in Fig.~\ref{fig:fitall}, along with the parameter values.  The fit reproduces the data quite well, both before and after perihelion, thus supporting the proposed scenario.

Before perihelion, the dependence of ${\rm Q}({\rm FeI})$ on heliocentric distance is consistent with thermal equilibrium, as found in Paper~I. The temperature dependence of Ni(CO)$_4$ as a function of the heliocentric distance is shallower. Post-perihelion, the temperature profiles are flatter for both Fe(CO)$_5$ and Ni(CO)$_4$.  In general, steeper (shallower) Q($r_{\rm h}$) values are obtained using steeper (shallower) $T (r_{\rm h})$ relations.

We stress that the active areas of Fe(CO)$_5$ and Ni(CO)$_4$ should be interpreted with caution, as they depend on the abundance ratio of Ni(CO)$_4$ over Fe(CO)$_5$ in the ice phase. This ratio may be close to the cosmic (solar) Ni/Fe abundance ratio, $\log_{10}$(Ni/Fe) = $-1.25$ \citep{Asplund2009}. Additionally, active areas and temperatures are degenerate in the model, different parameter combinations providing equally good fits.

\begin{figure}[]
\centering
\resizebox{0.95\hsize}{!}{\includegraphics*{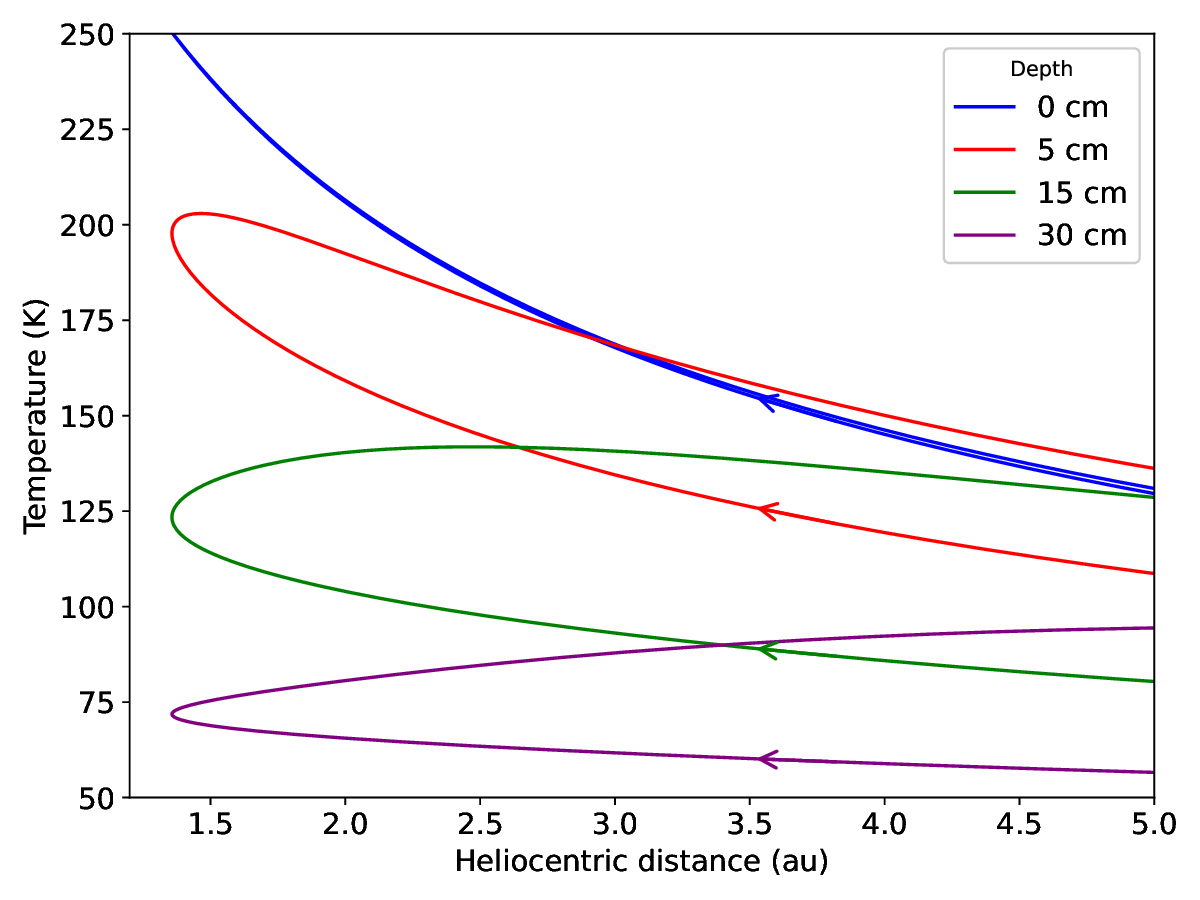}}
\caption{Temperature profiles of comet 3I as a function of the heliocentric distance. These profiles were computed with the thermal model of \citet{Yaginuma2026}, using their public code. An albedo of 0.1 and a thermal conductivity of 10$^{-3}$ W m$^{-1}$ K$^{-1}$ were used. The arrow indicates the direction of motion, from pre- to post-perihelion. The temperature profile at the surface corresponds to $T \simeq 300 \; r_h^{-1/2}$.}
\label{fig:Tprofile}
\end{figure}

\begin{figure*}[]
\centering
\resizebox{0.95\hsize}{!}{\includegraphics*{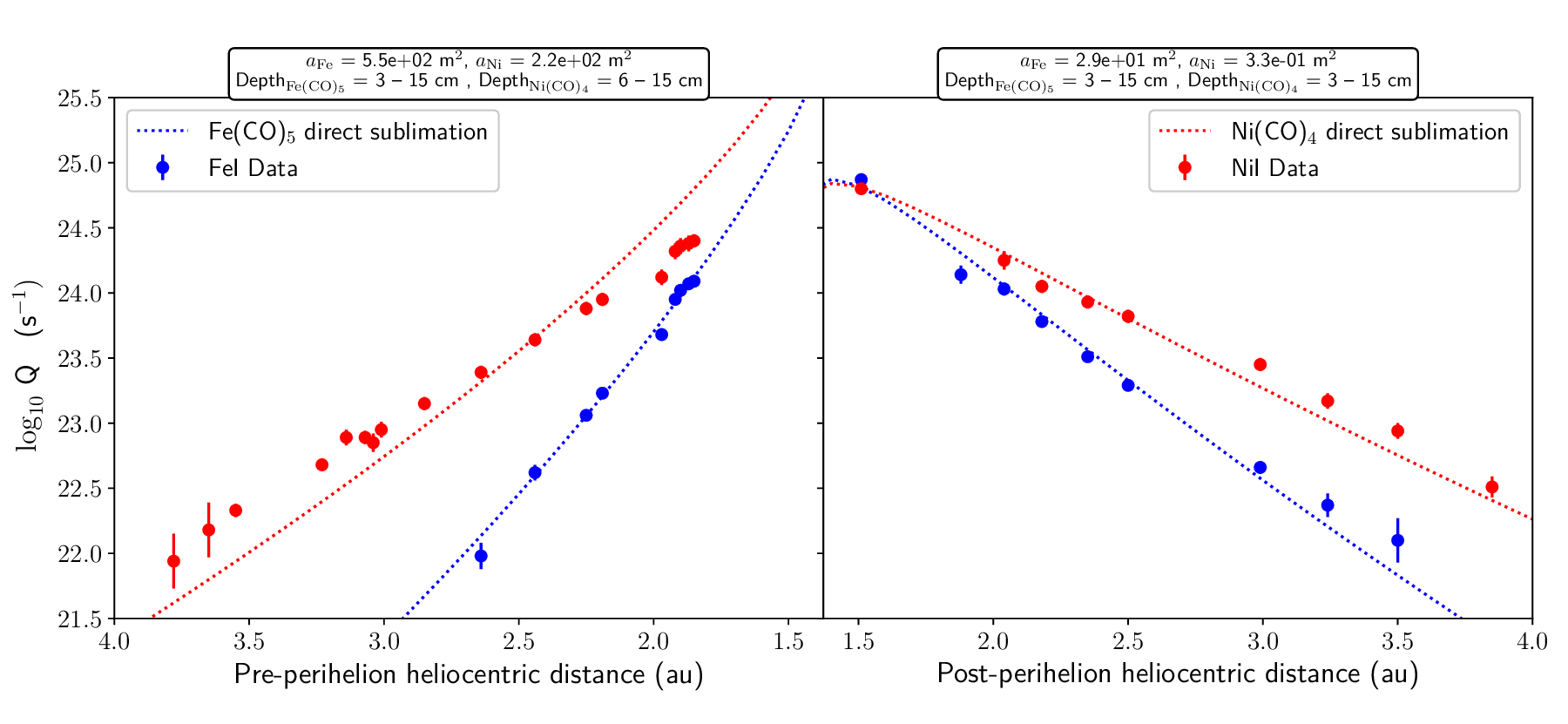}}
\caption{Same as Fig.~\ref{fig:fitall} but the model incorporates temperature profiles computed at various depths within the nucleus according to the thermal model of \citet{Yaginuma2026}. Pre- and post-perihelion datasets were fitted independently. The fit parameters and the considered layer depths are given above each panel.}
\label{fig:fityagi}
\end{figure*}

\begin{figure*}[]
\centering
\resizebox{0.95\hsize}{!}{\includegraphics*{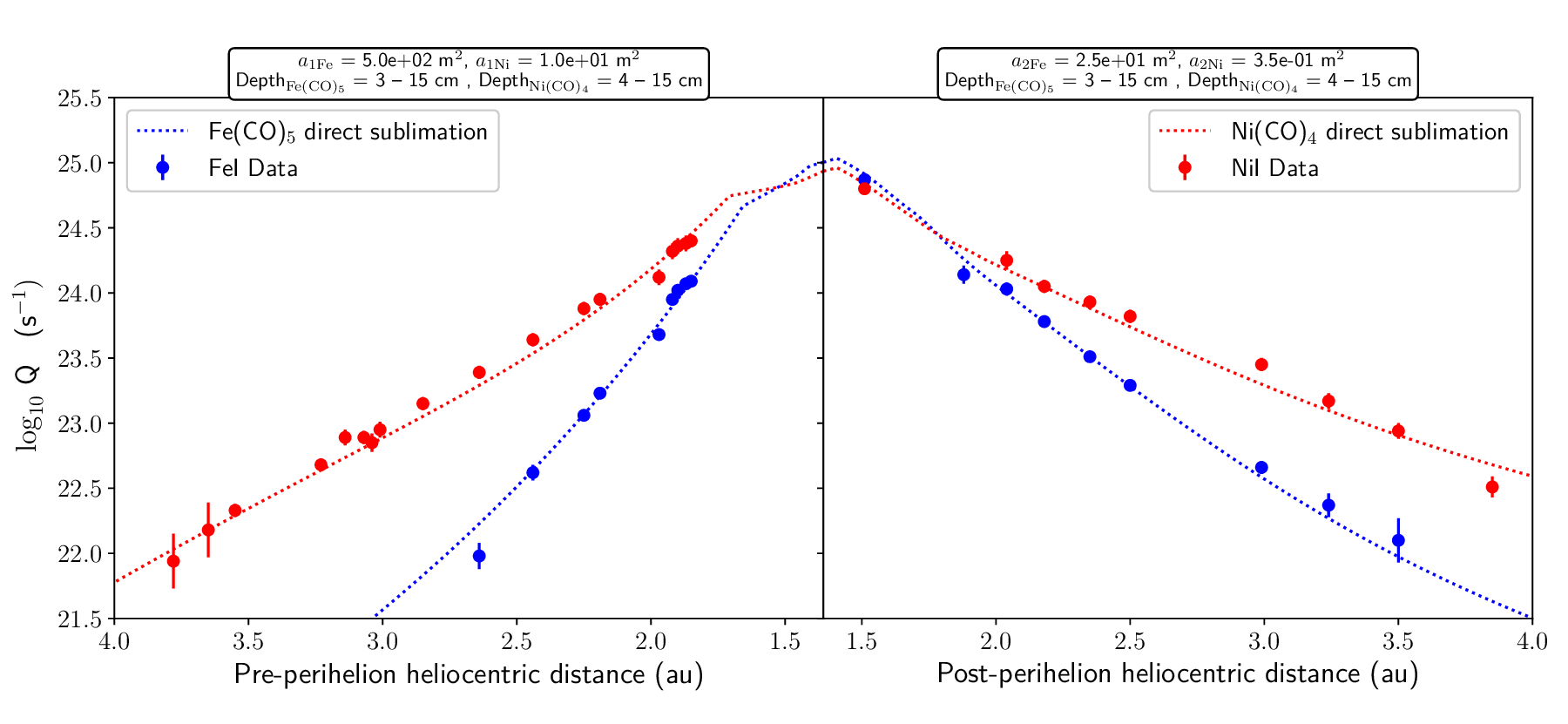}}
\caption{Same as Fig.~\ref{fig:fityagi} but with modifications of the thermal model to ensure the continuity of the production rates at the perihelion and to reproduce the NiI inbound leg (see text). }
\label{fig:fityagi2}
\end{figure*}

\subsection{Incorporating a thermal model of the nucleus}

The temperature profiles that reproduce the observations are roughly consistent with those computed by \citet{Yaginuma2026} using a detailed thermal model. Some of these profiles are shown in Fig.~\ref{fig:Tprofile}. Before perihelion, the Fe(CO)$_5$ sublimation could occur close to the cometary surface approximately in thermal equilibrium. Since flatter $T (r_{\rm h})$ relations are found below the surface, the pre-perihelion sublimation of  Ni(CO)$_4$ and the post-perihelion sublimation of both Ni(CO)$_4$ and Fe(CO)$_5$ could thus occur below the porous surface. To further examine this hypothesis, we incorporated the temperature profiles of \citet{Yaginuma2026} into our model, considering a range of depths below the surface.  We computed the sublimation rates for each layer, with a depth increment of one centimeter. The total sublimation rates were obtained by adding the contributions of the different layers, noting that the sum is dominated by the layers with the highest temperatures, which are usually closer to the surface. In this model, the areas releasing Ni(CO)$_4$ and Fe(CO)$_5$ likely depend on the surface area of the walls, which itself may depend on the layer depth.

In Fig.~\ref{fig:fityagi} we show the fits for the inbound and outbound legs independently. For NiI after perihelion and for FeI, the temperature profiles of \citet{Yaginuma2026} can reproduce the heliocentric distance dependence of the production rates, assuming that sublimation occurs deeper than a few centimeters. For NiI before perihelion, the data could not be accurately reproduced because the temperature profiles from the thermal model are still too steep at depths of up to 15 cm. Below that depth, sublimation rates are too low, requiring emitting areas larger than the nucleus area ($\sim$ 20 km$^2$, \citealt{Hui2026}) to fit the production rates. Furthermore, the thermal model fails to reproduce the combined dataset, as evidenced by the large discontinuity between the pre- and post-perihelion production rates. The production rates measured just after perihelion are at least one order of magnitude lower than the rates predicted by the thermal model just before perihelion. This indicates an abrupt decrease in active areas. To reproduce the data and ensure the continuity in the production rates at the perihelion, we made several ad hoc modifications to the temperature profiles of \citet{Yaginuma2026}. First, to reproduce the NiI inbound leg, we added a transient heat source of $T \sim 120$ K to the layers deeper than 5~cm. This additional source is simply modeled as $\Delta T ({\rm K})= 13 \times \exp[-((T-120)/30)^2]$. Although this transient temperature increase was added to all temperature profiles deeper than 5 cm, it only significantly affects the pre-perihelion behavior of Ni(CO)$_4$. Such a heat source could explain the extreme NiI/FeI abundance ratio observed in the early phases of the comet activity (Fig.~\ref{fig:c05}). Interestingly, the temperature of this additional heat source roughly corresponds to the temperature of the phase transition from amorphous to crystalline water ice \citep{Jenniskens1996,Prialnik2024}. Second, we limited the temperature of the nucleus to a maximum value of 200-210~K and introduced a smooth step function to model the change in the active areas of Ni(CO)$_4$ and Fe(CO)$_5$ through the perihelion: 
\[
a =
\begin{cases}
a_1, & {\rm for}\hspace{3mm} x < -x_i,\\[6pt]
\displaystyle
a_2 + \left( \frac{x_i-x}{2x_i} \right)^3 (a_1-a_2), & {\rm for}\hspace{3mm} -x_i \leq x \leq x_i,\\[12pt]
a_2, & {\rm for}\hspace{3mm} x > x_i .
\end{cases}
\]
Pre-perihelion, $x = - r_h + r_p$, and post-perihelion, $x = r_h - r_p$, with $r_p$ = 1.357 au. $a_1$ and $a_2$ are the constant active areas before and after the sharp decrease near perihelion. $x_i$ is fixed to 0.30 au for FeI and 0.35 au for NiI before perihelion, and to 0.70 au for both after the perihelion. $a_1$ and $a_2$ are derived from fitting the data. At perihelion, $x = 0$, and $a = (1/8) \; a_1 + 7/8 \;a_2$ for both legs, ensuring the continuity of the active areas. With these modifications, we were able to simultaneously fit both the pre- and post-perihelion datasets, maintaining continuity of the active areas,  temperatures, and production rates through the perihelion (Fig.~\ref{fig:fityagi2}). We emphasize that these modifications are purely ad hoc. However, the flattening of the temperature near the perihelion could be related to the presence of icy grains in the coma \citep{Yang2025,Belyakov2026}.  If these grains are particularly abundant at perihelion, we could speculate that they could decrease the amount of sunlight reaching the nucleus.  The drop in active area could be related to the large amount of material ejected close to perihelion that removed several meters of the nucleus outer layers \citep{Frincke2026}.

\subsection{Discussion}

Although our models are simplistic, the pre- and post-perihelion production rates of NiI and FeI can be reproduced by assuming the direct sublimation of the Ni(CO)$_4$ and Fe(CO)$_5$ carbonyls. In particular, the high NiI/FeI abundance ratio can be explained by the higher sublimation rate of Ni(CO)$_4$. The full dataset can be reproduced using the thermal model of \citet{Yaginuma2026}, provided that a transient heat source is added to reproduce the NiI pre-perihelion production rates. Close to perihelion, a temperature maximum and a significant reduction in the active areas must also be introduced. Within this modified thermal model, carbonyl sublimation occurs below the surface of the nucleus.

The possibility of carbonyl sublimation occurring below the surface could indicate that the comet has already been heated, resulting in the depletion of the most volatile species from the upper layers.  This interpretation is consistent with the behavior of methane reported by \citet{Belyakov2026}. However, the high porosity of the cometary surface layers necessary for in-depth sublimation, and the possibility that some water ice remained amorphous suggest that the previous heating was moderate and only affected the most volatile species.

Although carbonyls have been found in the interstellar medium \citep{Tielens1996}, they have not yet been detected in cometary comae, including in comet 67P \citep{Rubin2022}. Spectroscopic signatures are expected around 5 and 18 $\mu$m  \citep{Fiorini2016}, but their detection could be challenging given the expected low abundance of these species.  Therefore, alternative interpretations remain valuable and should be investigated, such as the release of NiI and FeI atoms into the gas phase from superheated metallic nanoparticles \citep[e.g.,][]{Ip1998,Velampatti2026}. Metal atoms could also be released from  organometallic complexes involving polycyclic aromatic hydrocarbons (\citealt{Klotz1996,Bromley2021,Manfroid2021}). However, in that scenario, one would not expect the production rates of NiI and FeI to differ in their dependence on $r_h$, since it is a photo-desorption process.

\section{Conclusions}
\label{sec:conclu}

High-resolution spectra of the interstellar comet 3I were obtained with the VLT+UVES after perihelion, at heliocentric distances ranging from 1.88 to 4.35 au. In this paper, we focused on the evolution of the FeI and NiI emission lines until their disappearance, which occurred around 3.5 and 4.35 au, respectively. These observations complement those obtained before perihelion and reported in Paper~I. With observations obtained regularly from 1.9 to 4.4 au, both inbound and outbound, this series of high-resolution spectra currently constitutes the most exhaustive dataset documenting the evolution of NiI and FeI emission for a single cometary object.

The new observations confirm the exceptionally high production rates of NiI and FeI in comet 3I compared to those in Solar System comets and the interstellar comet 2I. The initially extreme pre-perihelion NiI/FeI abundance ratio evolved toward values comparable to those of Solar System comets at 2~au and exhibited a weaker dependence on the heliocentric distance after perihelion. We found that the production rates are asymmetric about perihelion. At a given heliocentric distance, post-perihelion production rates are higher than pre-perihelion rates. They also declined more slowly with increasing heliocentric distance than pre-perihelion rates. The asymmetry is more pronounced for FeI.

To explain these observations, we built a model in which the FeI and NiI atoms originate from the photodissociation of gaseous Fe(CO)$_5$ and Ni(CO)$_4$ carbonyls, assumed to be released by direct sublimation from the icy nucleus. This model reproduces the measured production rates, both pre- and post-perihelion. The higher volatility of Ni(CO)$_4$ naturally explains the elevated NiI/FeI abundance ratio. Desorption of carbonyls from sublimating CO, CO$_2$ and H$_2$O ices is expected to be negligible. The temperature profiles needed to reproduce the observations were found to be shallower than the equilibrium $T \propto r_h^{-1/2}$ relation. The full dataset can also be reproduced using the temperature profiles from the thermal model of \citet{Yaginuma2026}, but only with significant empirical modifications to the model. A transient heat source was added to reproduce the pre-perihelion production rates of NiI, while a temperature maximum and a significant reduction in the active areas were introduced near perihelion.  Within this modified thermal model, carbonyl sublimation occurs below the surface of the nucleus.
The asymmetry of the temperature profiles with respect to perihelion explains the observed asymmetry in the NiI and FeI production rates. 
The fact that carbonyl sublimation could occur below the surface suggests that the comet has already been heated, resulting in the depletion of the most volatile species from the upper layers. Interestingly, the $r_h$-dependence of the metal production rates could be used to infer the thermal history of the nucleus thanks to their sensitivity to the temperature profiles, should the carbonyl scenario be validated.

Although carbonyls provide a self-consistent physical scenario for explaining the release of metal atoms at large heliocentric distances, they remain undetected in cometary comae. Alternative explanations, such as release from superheated metallic nanoparticles, remain viable and merit further observational and laboratory testing. To confirm the carbonyl hypothesis and clarify the origin of metals in 3I and other comets, future targeted mid-infrared spectroscopy (5–18 µm), deeper searches for carbonyl signatures, and combined thermal and compositional modeling are necessary.

\section*{Data availability}
The UVES observations are available from the ESO Archive under program 116.28NL.002 (PI: Opitom) as of February 2027.

\begin{acknowledgements}
We thank the referee, Michal Drahus, for constructive comments. DH thanks Steve Bromley for useful discussions. DH and EJ are Research Directors at the F.R.S-FNRS. JM is honorary Research Director at the F.R.S-FNRS. RD acknowledges support from grant \#361233 awarded by the Research Council of Finland to M. Granvik. Based on observations made with the ESO Very Large Telescope at the Paranal Observatory under program 116.28NL.002.
\end{acknowledgements}

\bibliographystyle{aa}
\bibliography{references}

\onecolumn
\begin{appendix}

\section{UVES post-perihelion spectrum of comet 3I/ATLAS}
\label{app:specnife}

\begin{figure*}[h]
\centering
\resizebox{0.9\hsize}{!}{\includegraphics*{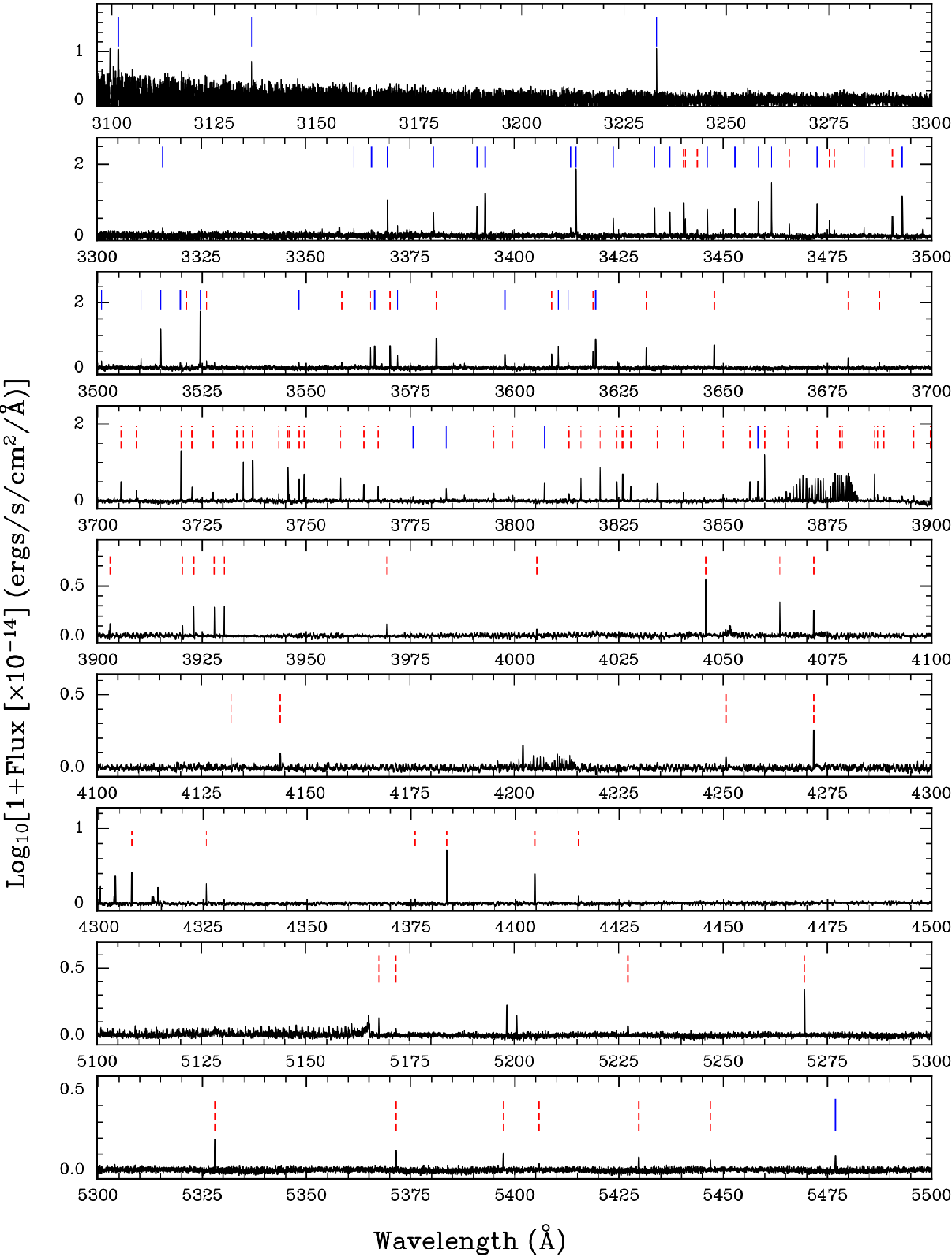}}
\caption{Combination of continuum-subtracted spectra of comet 3I obtained on December 6 and 10 with UVES. Many NiI (solid blue tick marks) and FeI (dashed red tick marks) emission lines are detected in this spectral range, though not between 4500~\AA\ and 5100~\AA.}  
\label{fig:spec_nife}
\end{figure*}


\section{Fluorescence model: Three-level versus multilevel}
\label{app:3levels}

\begin{figure}[h]
\centering
\resizebox{0.5\hsize}{!}{\includegraphics*{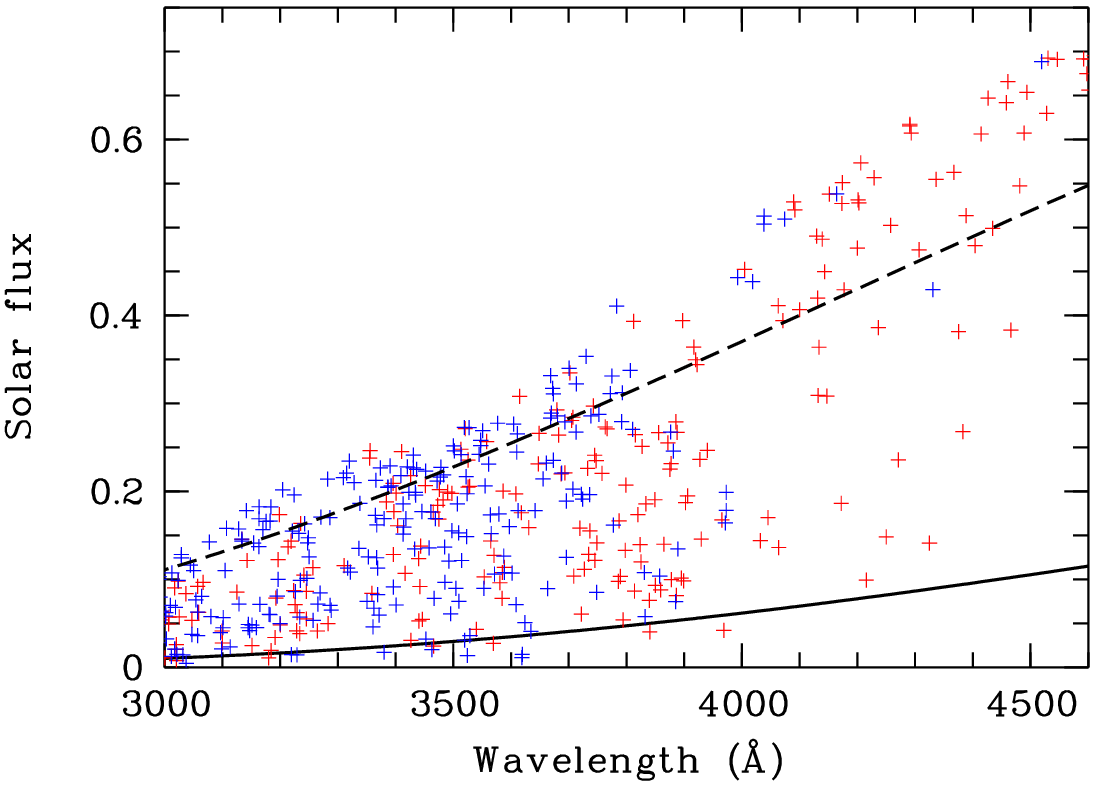}}
\caption{Solar flux (in arbitrary units) seen by the atomic NiI transitions (blue crosses) and the FeI transitions (red crosses), as used in the multilevel fluorescence model. The heliocentric distance of comet 3I and the Sun-comet relative velocity correspond to those on December 21, 2025. The solar flux approximated from blackbodies at temperatures of 4500~K (solid line) and 5800~K (dashed line) is also shown. The 4500~K blackbody clearly underestimates the true solar flux.}  
\label{fig:solarflux}
\end{figure}

\citet{Manfroid2021} showed that the simple three-level atomic model used by \citet{Preston1967} and \citet{Arpigny1979} yields NiI/FeI column density (abundance) ratios similar to those provided by the more realistic multilevel fluorescence model, particularly when the excitation temperature of NiI is increased by 180~K relative to the temperature derived from the analysis of the FeI lines. While the ratios are similar, the column densities themselves, and consequently the production rates, are not; the latter can be overestimated by as much as one order of magnitude. For example, using the three-level atomic model as detailed in \citet{Manfroid2021}, we found a FeI excitation temperature of approximately 4500~K for the dataset obtained for comet 3I on December 21 with setting 348. This temperature is typical of that found in Solar System comets  \citep{Manfroid2021}. In this framework, we obtained $\log_{10}$ Q(FeI) = 24.06, $\log_{10}$ Q(NiI) = 24.41, and $\log_{10}$ [Q(NiI)/Q(FeI)] = 0.35. While the NiI/FeI abundance ratio agrees with the value obtained from the multilevel model  within the uncertainties (Table~\ref{tab:data}), the production rates are about 0.5~dex too large. As shown in Fig.~\ref{fig:solarflux}, this discrepancy can be explained by the fact that the blackbody used in the three-level model underestimates the true solar flux seen by the NiI and FeI atomic transitions. Therefore, for the same observed emission line intensities, more atoms are needed to compensate for the fewer solar photons.

\section{Comparison with  Bromley's fluorescence model}
\label{app:sbcompa}

\begin{figure}[h]
\centering
\resizebox{0.9\hsize}{!}{\includegraphics*{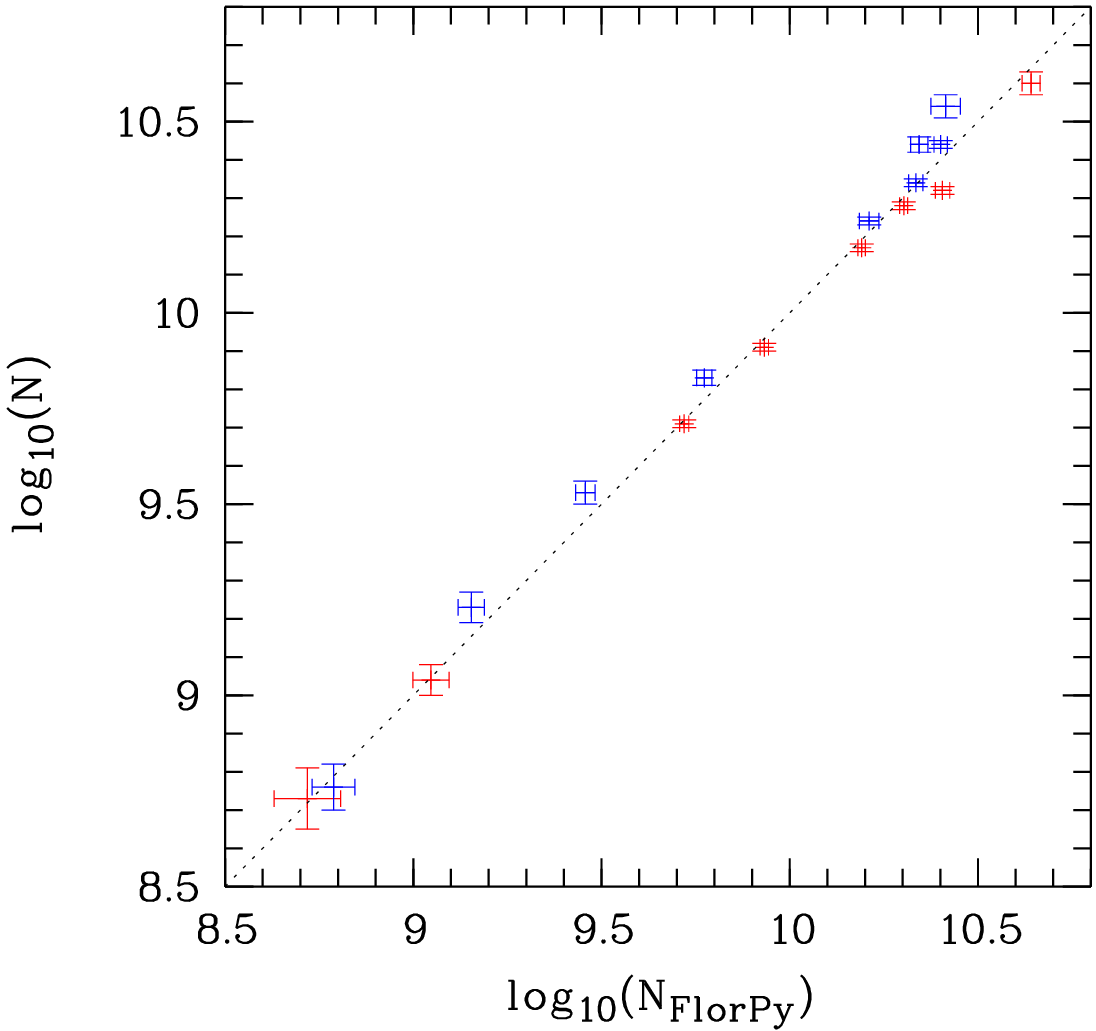}\hspace{4cm}\includegraphics*{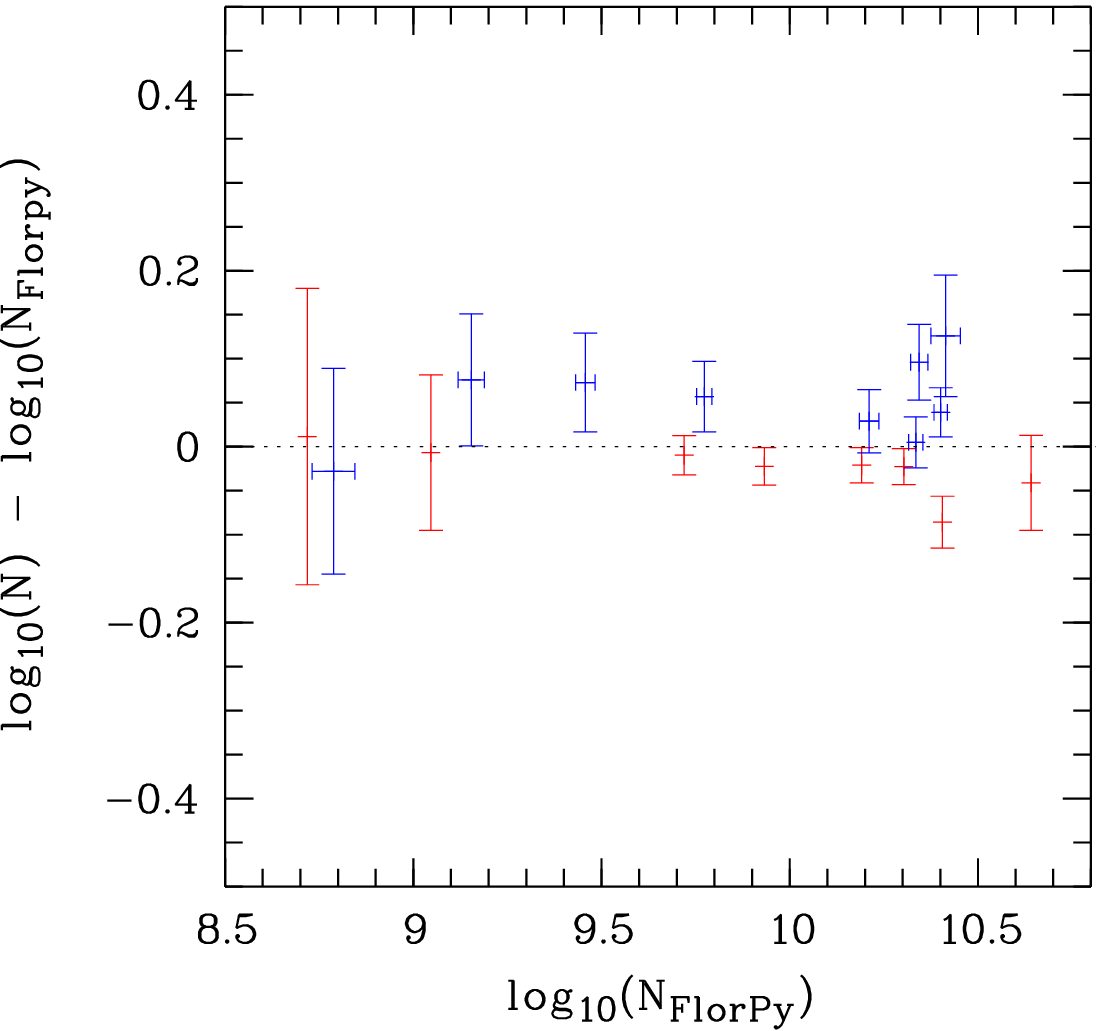}}
\caption{Comparison of the average column densities ($N$ in cm$^{-2}$) computed with our fluorescence model for each post-perihelion epoch to those obtained with Steven Bromley's model FlorPy \citep{Bromley2021,Bromley2026}. Our atomic data are taken from the Atomic Line List \citep{Vanhoof2018}, while FlorPy uses the NIST Atomic Spectra Database \citep{Kramida2026}. The column densities of NiI are shown in blue and those of FeI in red.}
\label{fig:sbcompa}
\end{figure}

\section{Additional table}
\label{app:tables}

\begin{table*}[h]
\caption{Revised pre-perihelion FeI and NiI production rates, and their ratio.}
\label{tab:data1}
\centering
\begin{tabular}{lccccccccccc}
\hline\hline
 Date & $r_h$ & n$_{\rm lines}$  & log$_{10}$ Q(FeI) & log$_{10}$ Q(NiI) & log$_{10}$ [Q(NiI)/Q(FeI)] & Instrument \\
 yyyy-mm-dd & au     &  FeI / NiI  &   s$^{-1}$     &   s$^{-1}$ &        &     \\
\hline 
2025-07-23  & 3.78 &    - /  2   &   -                & 21.94 $\pm$ 0.21  & -                & X-SHOOTER   \\
2025-07-27  & 3.65 &    - /  2   &   -                & 22.18 $\pm$ 0.21  & -                & X-SHOOTER   \\
2025-07-30  & 3.55 &    - /  4   &   -                & 22.33 $\pm$ 0.05  & -                & X-SHOOTER   \\
2025-08-09  & 3.23 &    - /  3   &   -                & 22.68 $\pm$ 0.05  & -                & X-SHOOTER   \\
2025-08-12  & 3.14 &    - / 20   & $<$ 21.76          & 22.89 $\pm$ 0.06  &   $>$ 1.13       & UVES \\
2025-08-14  & 3.07 &    - / 11   &   -                & 22.89 $\pm$ 0.05  & -                & X-SHOOTER   \\
2025-08-15  & 3.04 &    - / 14   & $<$ 22.07          & 22.85 $\pm$ 0.07  &   $>$ 0.78       & UVES \\
2025-08-16  & 3.01 &    - / 12   &   -                & 22.95 $\pm$ 0.06  & -                & X-SHOOTER   \\
2025-08-21  & 2.85 &    - / 14   &   -                & 23.15 $\pm$ 0.05  & -                & X-SHOOTER   \\
2025-08-28  & 2.64 &    5 / 27   &  21.98 $\pm$ 0.09  & 23.39 $\pm$ 0.05  & 1.41 $\pm$ 0.11  & UVES \\
2025-09-3/4 & 2.44 &   17 / 35   &  22.62 $\pm$ 0.06  & 23.64 $\pm$ 0.05  & 1.03 $\pm$ 0.08  & UVES \\
2025-09-10  & 2.25 &   40 / 35   &  23.06 $\pm$ 0.03  & 23.88 $\pm$ 0.05  & 0.82 $\pm$ 0.06  & UVES \\
2025-09-12  & 2.19 &   46 / 36   &  23.23 $\pm$ 0.03  & 23.95 $\pm$ 0.05  & 0.72 $\pm$ 0.06  & UVES \\
2025-09-20  & 1.97 &   11 / 18   &  23.68 $\pm$ 0.04  & 24.12 $\pm$ 0.06  & 0.44 $\pm$ 0.07  & X-SHOOTER   \\
2025-09-22  & 1.92 &   16 / 18   &  23.95 $\pm$ 0.04  & 24.32 $\pm$ 0.06  & 0.37 $\pm$ 0.07  & X-SHOOTER   \\
2025-09-23  & 1.90 &   21 / 18   &  24.02 $\pm$ 0.03  & 24.36 $\pm$ 0.06  & 0.35 $\pm$ 0.07  & X-SHOOTER   \\
2025-09-24  & 1.87 &   18 / 20   &  24.07 $\pm$ 0.04  & 24.38 $\pm$ 0.06  & 0.30 $\pm$ 0.07  & X-SHOOTER   \\
2025-09-25  & 1.85 &   16 / 20   &  24.09 $\pm$ 0.03  & 24.40 $\pm$ 0.05  & 0.31 $\pm$ 0.07  & X-SHOOTER   \\
\hline
\end{tabular}
\end{table*}

\section{Comparison with other measurements}
\label{app:datacompa}

\citet{Belyakov2026} discovered a forbidden NiI line at 7.507 $\mu$m in the spectrum of comet 3I obtained on December 16, 2025, using the James Webb Space Telescope (JWST) mid-infrared medium-resolution spectrometer (MIRI). Assuming fluorescence emission and direct ejection from the nucleus, they measured log$_{10}$ Q(NiI) = 23.89$\pm$ 0.15, with Q in s$^{-1}$. They also uncovered a fainter NiI line at 11.307 $\mu$m, from which they derived log$_{10}$ Q(NiI) = 24.2$\pm$0.2, under the same hypotheses. These independent measurements are in excellent agreement with our value of log$_{10}$ Q(NiI) = 24.05$\pm$0.05, obtained on December 15, 2025. These results support the hypothesis that the NiI forbidden lines at 7.507 and 11.307 $\mu$m are primarily due to fluorescence. 

\citet{Roth2026c} reported the detection of two NiI lines at 3.119 and 3.915~$\mu$m on December 23, 2025, using the JWST NIRSpec integral field unit. They derived log$_{10}$ Q(NiI) in the range 24.3$-$24.9, with different assumptions about the parent. They assumed an expansion velocity of 0.31 km~s$^{-1}$, which is lower than the velocity of 0.54 km~s$^{-1}$ used by \citet{Belyakov2026}. Rescaling to the higher velocity would give log$_{10}$ Q(NiI) in the range 24.5$-$25.1, which is higher than the production rates measured by \citet{Belyakov2026} on December 16, and the value measured with UVES on December 21, log$_{10}$ Q(NiI) = 23.93$\pm$0.05 (Table~\ref{tab:data}).

\citet{Zhao2026} obtained multi-epoch low-resolution ($R \sim 300$) spectra of comet 3I from December 2, 2025, to January 20, 2026, using the Xinglong 2.16m and the Lijiang 2.4m telescopes. The spectra covered the spectral ranges 3300-6600 \AA\ and 3200-9300 \AA, respectively. During this period, the comet was at heliocentric distances ranging from 1.8 to 3.3~au. The production rates they derived for FeI are in excellent agreement with ours, while their NiI production rates are about a factor of two higher, likely due to an overcorrection in the flux calibration at the blue end of the spectrum \citep{Zhao2026}. The variation of the production rates of both FeI and NiI with heliocentric distance was found to be shallower post-perihelion than pre-perihelion, as confirmed by our observations (Fig.~\ref{fig:q01}).

\citet{Hoogendam2026} reported post-perihelion observations of comet 3I obtained with the Keck Cosmic Web Imager on November 16, 2025, when the comet was at $r_h$ = 1.51~au, thus closer to perihelion than our UVES observations. With a resolving power of $R \sim 1800$, several NiI and FeI lines were detected in the spectral range 3300-5500 \AA. Production rates were computed using both a simplified three-level model and a multilevel fluorescence model.  Since production rates computed with the simplified model are likely overestimated (Appendix~\ref{app:3levels}), we only considered those obtained with the multilevel model: log$_{10}$ Q(FeI) = 24.87$\pm$0.02 and log$_{10}$ Q(NiI) = 24.80$\pm$0.01 (N.B. The NiI and FeI production rates were inadvertently flipped in their paper; Hoogendam, private communication). These values are in agreement with the $r_h$-dependence of our measurements extrapolated at $r_h$ = 1.51~au (Figs.~\ref{fig:q01}--\ref{fig:c05}).

\citet{Mura2026} obtained long-slit spectra of comet 3I with the 1.22m telescope of the Asiago Observatory, when the comet was at $r_h$ = 2.17~au. The spectra covered the wavelength range 3300-4570 \AA\ with a resolving power of $R \sim 2000$. Several 
NiI and FeI lines were detected from which log$_{10}$ Q(NiI) = 24.05$\pm$0.07 and log$_{10}$ Q(FeI) = 23.85$\pm$0.07 were estimated. These values are in excellent agreement with the values we obtained at $r_h$ = 2.18~au (Table~\ref{tab:data}).

\end{appendix}

\end{document}